\documentclass[12pt]{article}
\usepackage{graphicx}
\usepackage{cancel}[]
\usepackage{amssymb,amsmath}
\usepackage{ifpdf}
\usepackage{hyperref}[]
\usepackage{fullpage}
\usepackage{color}
\usepackage{ulem}
\usepackage{verbatim}
\usepackage{float}
\ifpdf
    \DeclareGraphicsExtensions{.pdf}
\else
    \DeclareGraphicsExtensions{.eps}
\fi

\usepackage{graphicx,epsfig}
\usepackage{cancel}[]
\usepackage{amssymb,amsmath}
\usepackage{ifpdf}
\usepackage{hyperref}[]
\def\lsim{<\kern-2.5ex\lower0.85ex\hbox{$\approx$}\ }
\def\rsim{>\kern-2.5ex\lower0.85ex\hbox{$\approx$}\ }

\def\LAMBDABAR
{\hbox{$\lambda$\kern-0.52em\raise+0.45ex\hbox{--}\kern+0.2em}}

    \DeclareGraphicsExtensions{.eps}

\begin{document}

 \centerline{\Large\bf{Possible scheme for observing }}
\centerline{\Large\bf{acceleration (Unruh) radiation }}
\vspace{.20in}

 \centerline{A. C. Melissinos }
\centerline{\it{Department of Physics and Astronomy, University of Rochester }}
\centerline{\it{Rochester, NY 14627-0171, USA}}
\vspace{.20 in}

\section{Introduction}

What is acceleration radiation and why is it interesting to observe it? 

Following Hawking's papers on the evaporation of black holes \cite{Hawking}, it was soon
realized that by the correspondence principle, something analogous must hold for an accelerated
observer. Unruh and Davies \cite{Unruh, Davies} pointed out that an accelerated observer in his own frame of reference sees an altered vacuum, and this should lead to a virtual thermal 
radiation spectrum at temperature
\begin{equation} T=\frac{\hbar a_p}{2 \pi c k_B} \end{equation}
where $a_p$ is the {\it{proper}} acceleration of the observer.

In principle there is no new Physics, except that since acceleration is equivalent to a gravitational field, 
it was speculated that observation of acceleration radiation could elucidate issues in gravitation and in
particular the crossing of a black hole horizon. Forty years after the Unruh publication there has been no 
explicit (experimental) observation of acceleration radiation in spite of many proposals \cite{Rogers, Matsas}, and
analogies to other systems where shock waves appear \cite{Various}. One concludes that acceleration radiation is nothing new, yet it would be interesting to observe it in the electromagnetic domain. The difficulty is
that the radiation is very weak, and as Eq(1) shows, a temperature of 1 K is reached for an acceleration 
$a_p = 3\times 10^{20}\ {\rm{m/s^2}} \approx 3\times 10^{19}{\rm{g_{\odot}}}$. Furthermore, accelerated 
electric charges radiate profusely and this can mask the acceleration radiation. For instance, for an
electric charge in a circular orbit, the acceleration radiation becomes equal to the Larmor (synchrotron) 
radiation for acceleration $a_p \approx 3\times 10^{31}\ {\rm{m/s^2}}$; this corresponds to $T \approx 10^{10} \ ^{\circ}K$. 

Linear acceleration of the order mentioned above is difficult to generate, while high acceleration is reached in circular motion especially of relativistic particles. Another difference is that (constant)
linear acceleration leads to hyperbolic motion \cite{Rindler}, which lies outside the light cone and is confined within its own cone (or ``Rindler" wedge). This raises the question whether acceleration radiation is restricted to linear acceleration or is also present for circular motion. A careful analysis of the polarization of accelerated electrons in the LEP ring \cite{Bell} failed to show exact agreement with the temperature predicted by the acceleration radiation, but instead agrees exactly with the QED prediction \cite{Jackson1}. 

The ``Unruh" photons are, of course, virtual \cite{Holstein}. To detect them we must indirectly observe an effect that is dependent on the presence of the thermal bath; this is the idea proposed in \cite{Bell}, and by Rogers \cite{Rogers}. An alternate approach is to consider the scattering of the virtual 
(Unruh) photons, for instance off electrons, by which process the photons are placed on the mass shell and can be directly detected. Such conditions can be realized in a free-electron-laser and we consider the case of the LCLS (Linac Coherent Light Source) at Stanford \cite{LCLS1, LCLS2}. 
An important aspect of the presently operating  FEL's is that the electron beam becomes microbunched. This 
enhances the signal of the scattered Unruh photons, which nevertheless is difficult to detect in the presence of the copious spontaneous radiation in the same energy range.

\section{LCLS parameters}

We use the specifications from the 1998 LCLS study \cite{LCLS1} and the 2010 first lasing report \cite{LCLS2}. For the beam energy we use

$E = 14 $ GeV

and for the undulator mean field

$ B = 1.4$ T

These values lead to proper acceleration
\begin{equation} a_p = \frac{e}{m}B\gamma\beta c = 2.1 \times 10^{24}\ {\rm{m/s^2}} 
\end{equation}
and thus to an ``Unruh temperature" $$ T = 8,560\ {\rm{K}}\qquad{\rm{or}}\quad k_{B} T= 0.74 \ {\rm{eV}}$$

The virtual (Unruh) photon number density, per unit energy, in the electron rest frame is 
\begin{equation} \frac{dn}{\hbar d\omega} = 
\frac{1}{\pi^2 (\hbar c)^3} \frac{(\hbar\omega)^2}{e^{\hbar \omega/k_b T}-1} \end{equation}

The virtual photons are distributed isotropically in the electron rest frame and scatter off the 
(comoving) electrons 
in the beam. This changes their angular orientation (which remains isotropic), but not their energy;  virtual photons that have scattered off the beam are now real, isotropically distributed and in the electron rest frame have an energy distribution given by the Planck law of Eq(3), where for the scattering probability we use the Thomson cross section, $\sigma_T = (8\pi/3)r_0^2$. In the lab frame the photon energy is \begin{equation} E = \overline{\gamma}
\omega_0 (1+\overline{\beta}\cos\theta)\end{equation} with $\overline{\gamma}$  the Lorentz factor of the electron's {\it{forward}} motion and $\theta$ is the direction of  the photon ($\theta =0$ in the forward direction) in the electron rest frame. We can safely set $\overline{\beta} = 1$.
In Eq(4) $\overline{\gamma}$ differs from $\gamma=E/mc^2$ by the so-called K-factor that depends on the strength of the undulator, and for the LCLS,
$K=(e/m)(B\lambda_0/c)= 3.71$. We have $$ \overline{\gamma} = \gamma/\sqrt{1 + K^2/2}$$
For the LCLS parameters, the scattered (Unruh) photon spectrum in the lab frame is shown in Fig.1 
and peaks at $\sim 10.4\ {\rm{keV}}$.\\

\newpage
To calculate the rate of scattered photons we use the following parameters from the LCLS report \cite{LCLS1}

Number of bunches per second \hspace{10pt} $ n_b = 120.$

Number of microbunches per bunch \hspace{10pt}  $n_{\mu b}$ = 60.

Electrons/bunch (1 nC in each bunch)\footnote{This assignment leads to an average current of 120 nA, instead of the quoted average current of 72 nA.} \hspace{10pt} $N_b = 6\times 10^9.$      

Electrons/microbunch  \hspace{10pt} $N_{\mu b} =  10^8 .$

Undulator length \hspace{20pt}  $L_u = 10^4\ {\rm{cm}}.$

Forward Lorentz factor \hspace{10pt} $\overline{\gamma} = \gamma /2.81 \approx 10^4.$
\\

Due to the well known FEL instability \cite{Pellegrini} the electron current is microbunched and at half saturation the microbunch length (in the lab frame) is $\Delta t^{\mu b}_{L} \approx 0.2\ {\rm{fs}} = 2\times 10^{-16}\ {\rm{s}}$. The cloud of virtual photons in the electron rest frame persists for a time interval $$ \Delta t^{v}_{R} = \hbar/\delta E \approx 6\times 10^{-16}({\rm{eV}})
/0.7 ({\rm{eV}}) \approx 1\ {\rm{fs}}.$$ 
This is to be compared with the duration of the microbunch passage as seen in the electron rest frame
$\Delta t^{\mu b}_{R} = \overline{\gamma} \Delta t^{\mu b}_{L} \approx 2\times 10^{-12}\ {\rm{s}}$. Thus only a fraction
$$ R= \Delta t^{v}/\Delta t^{\mu b} \approx 5\times 10^{-4}$$
of the electrons in the microbunch will scatter coherently from the Unruh photons\footnote{Assuming full transverse coherence.}. The scattering rate for Unruh photons in a single microbunch is $(R N_{\mu b})^2$ times that for a single electron.

{
{
{

In transforming the scattered photon spectrum (i.e. the distribution in energy), we pick the (rest frame) emission angle randomly and integrate over all events in a given lab energy bin. Of course the lab angle and energy are correlated, but due to the Lorentz boost almost all photons are contained in a forward cone of angular range 
$\Theta \sim 1/\overline{\gamma}$. The resulting spectrum in the lab frame is shown in Fig.1 for the LCLS parameters specified above. The peak of the distribution is at $E_L = 10.4$\ keV, and the Unruh photon rate at the peak is \begin{equation}dn/dE_L = 6 \times 10^{4}\ {\rm{ photons/s/keV}}. \end{equation}

The peak of the Unruh radiation lies above the end point of the first harmonic (8.3 keV).
The spontaneous emission in this energy range comes from all higher harmonics. In Fig.2 we show the contribution from the second harmonic as calculated according to \cite{Jackson2}, for a 
bandwith of .03$\%$ (adopted in \cite{LCLS1}), so that at $E=10.4$\ keV,

\begin{equation}dn/dE|_{spont} =2\times 10^{10}\ {\rm{photons/s}}/(3\times 10^{-4} E) \approx
 7\times 10^{12}\ {\rm{photons/s/keV}}. \end{equation} This rate is eight orders of magnitude higher than 
the peak Unruh rate precluding the direct identification of the contribution of the Unruh radiation.\\

\section{Extreme acceleration}

Recently C. Pellegrini and D. Reis, have been considering experiments in Strong
Field QED (SFQED) at the LCLS \cite{Reis}. This would involve reversing the
direction of the 1A x-ray beam by back-scattering it of suitable crystals,
and after focusing the back-scattered beam to a 10 nm spot, bringing 
it into collision with the 14 GeV electron beam. Accounting for the
highly relativistic motion, the electrons in their rest frame see an
Electric field \begin{equation} E_r \approx 2 \gamma_e E_x \approx 1.3
\times 10^{18}\ {\rm{V/cm}} \approx 100 E_c \end{equation}
where $E_c$ is the (Schwinger) critical field and where QED is in the 
non-pertubative, strong field, regime. Correspondingly the electrons in their 
rest frame are subject
to a (proper) acceleration \begin{equation} a_p = 2\times 10^{31}\ {\rm{m/s^2}}
\end{equation} To reach this field it is assumed that x-ray  power of 
1 TW (10 mJ in $\Delta t = 10$ fs) can be focused to a 10 nm spot,
resulting in an intensity $I =10^{24}\ {\rm{W/cm^2}}$ in the laboratory
frame, and thus $<E_{L}> = 2\times 10^{13}\ {\rm{V/cm}}$. Given that 
the electron's Lorentz factor is $\gamma = 3\times 10^4$ and that the 
x-rays collide head-on with the electrons, the field in the electron rest frame 
reaches $\approx 100 E_c$.

Under these assumptions the Unruh temperature is 
\begin{equation} T=\frac{\hbar a_p}{2 \pi c k_B} = 8.6\times 10^{10}\
 ^{\circ}K,\end{equation}
which corresponds to an energy of $\approx 7 $ MeV, at the peak of the
Unruh (Planck) spectrum. This poses an important paradox if we follow the 
arguments used in the previous sections: how can an electron in its rest frame
emit a 7 MeV gamma-ray, even if subject to extreme acceleration  by
an external electric field? 

The difficulty arises because the virtual Unruh photons were treated as real.
The approach that we used earlier was that the virtual photons Thomson scatter from
the electrons and this places them on the mass shell. In this process the
virtual and ``real" photons have the same energy. This assumption 
is not valid for the high energy (virtual) photons, which can loose significant
energy when scattering from stationary electrons. The photon recoil energy, $\omega '$, depends
on the change in the photon's direction as in simple Compton scattering,
\begin{equation}\omega ' =\frac{m}{m/\omega + (1-{\rm{cos}}\theta)}\end{equation}
where $\theta$ is the scattering angle from the direction of the incident 
photon. When $m/\omega \ll 1$, as in the present case, the scattered photon
energy never exceeds the electron mass, $m$. However the direction of the photon,
 after scattering (now assumed real) 
in the electron rest frame determines the photon energy when transformed back
into the lab system. 

Eq.(10) correlates the energy of the scattered photon (by now assumed 
real) with its (virtual) energy in the electron rest frame. This is not sufficient 
to obtain the energy of the scattered photon in the lab frame because
we need to know its emission angle in the rest frame; the latter depends
not only on the scattering angle $\theta$ but also on the  
direction of incidence (before scattering) of the virtual photon. It is interesting  
that when the incoming photon is directed in the backward direction 
in the electron rest frame, the photon energy in the lab frame never 
exceeds the incident (beam) electron energy. This is shown in Fig.3 
which gives the (Unruh) photon energy in the lab frame (y-axis) as
a function of the photon energy in the rest frame (x-axis), integrated 
over all scattering angles, but for backward incidence. 
Since only this choice of virtual photon incidence 
does not violate our (naive) concept of energy conservation, we adopt 
it for all further estimates. 

In evaluating the rate of scattered virtual photons we use the Klein-Nishina 
cross section, and the interaction length equal to twice the Rayleigh length
of the x-ray beam (estimated in the lab frame to be)
$l_{int}^{lab} = 2z_R =  2\times 10 ^{-4} \ {\rm{cm}}.$
The spectrum of observable Unruh photons, in the lab frame, and under the above
assumption of backward incidence, integrated over all scattering angles 
is shown in Fig.4. The number of scattered photons per MeV per electron, is given as a function
of the lab photon energy per incident electron. Using the Klein-Nishina cross
section extends the spectrum
slightly towards lower energies as compared to the case when the Thomson cross section 
is used.
If these assumptions are correct, the manifestation of Unruh radiation would
be an accumulation of photons at the primary electron energy, and within the 
$\gamma \Theta \approx 1$ cone.\\

Integrating the spectrum of Fig.4 over energies, yields approximately 
one Unruh photon per incident electron, the energy of the Unruh photons being 
almost equal to the electron energy.  
Numerically, this high yield can be traced to the 
high Unruh temperature, $ T\sim 10^{11} \ ^{\circ}K$  which results in 
a very high density of virtual photons\footnote{Such temperature 
is reached in the first $1/10^{th}$ of a second after the ``big bang".}. \\

{\bf{Backgrounds:}} There are two sources of background, one is the hard photons 
from the scattering of the x-ray beam off the electrons, and the other is 
the photons radiated by the electrons as they are subject to the extreme acceleration
imposed by the Electric field. Given the electron energy and the high x-ray 
energy (10 keV), the spectrum of the back-scattered photons reaches the
kinematic limit, i.e the electron beam energy. Thus it is not possible to
differentiate them from the Unruh photons on kinematic grounds. 

To calculate the scattering probability (per electron) we let
\begin{equation} W_1 = \rho_{\omega} \sigma_C (1+\beta {\rm{cos}}\alpha) l_{int}^{lab}, \end{equation}
where $\beta {\rm{cos}}\alpha \approx 1$ with $\alpha$ the crossing angle, and $l_{int}^{lab}$ is the 
length of the interaction region as used previously.  The photon density in the
incoming beam is $$\rho_{\omega} \approx 10^{31}\ {\rm{cm^{-3}}}$$ 
and since $l_{int}^{lab} = 2z_R =2\times 10^{-4}$ cm,
we find, $W_1 \approx 2\times 10^{2}$. This is not surprising given the high x-ray density
density at the focus, but indicates that the electron beam will be strongly attenuated as it
approaches the focal region

Finally we examine the Larmor radiation of the accelerated electrons. There is
no simple (textbook) estimate of the expected frequency spectrum because there
is no apparent circular motion. To estimate the critical frequency we
can proceed as follows. In circular motion
\begin{equation} \omega_c = \frac{3}{2} \gamma^{3} \left(\frac{c}{\rho}\right)=
\frac{3}{2}\gamma \frac{a_p}{c} \end{equation} where we used the fact that the 
proper acceleration $a_p = \gamma^2 c^2/\rho$. It follows that the critical frequency 
in the electron rest frame  is \begin{equation} \omega_c = 3\times 10^{23}\ {\rm{r/s}} 
\qquad {\rm{and}} \qquad  \hbar \omega_c = 200\ {\rm{MeV}} \end{equation}
Thus in the lab frame the Larmor spectrum extends all the way to the kinematic
limit, i.e. the electron energy.   

As to the power radiated due to the Larmor process we use the standard expression
in the electron rest frame
 \begin{equation}P =\frac{2}{3} \frac{e^2}{4 \pi \epsilon_0} \frac{1}{c^3} a_p^2
\approx 2\times 10^9\ {\rm{W}} \end{equation}
As before we estimate the time that the electron spends (in its own rest frame) 
in the focal region  to be
$\tau = l_{int}^{rest}/c = 10^{-19}$ s, yielding a radiated energy of \begin{equation}
U_{Larmor} = P\tau \approx 1\ {\rm{GeV}} \end{equation}
per electron. This appears to be less than the energy carried by the Unruh photons.\\

{\bf{Caveat:}} In this section it was assumed that the electrons would reach the region 
where the x-ray beam is focused. This is debatable because of the early production of
$e^{+}e^{-}$ pairs and the development of an electromagnetic cascade which  would deplete 
the power in the incident x-ray beam \cite{Fedotov}, as also suggested by the above estimates
of the background.\\

{\bf{\large{Acknowledgement}}}

I thank Profs. K.T. McDonald and David Reis for pointing out an error in the first 
version of this note, and for extended discussions. 


\begin{figure}[H]
\centering
\includegraphics[width=130mm,height=80mm]{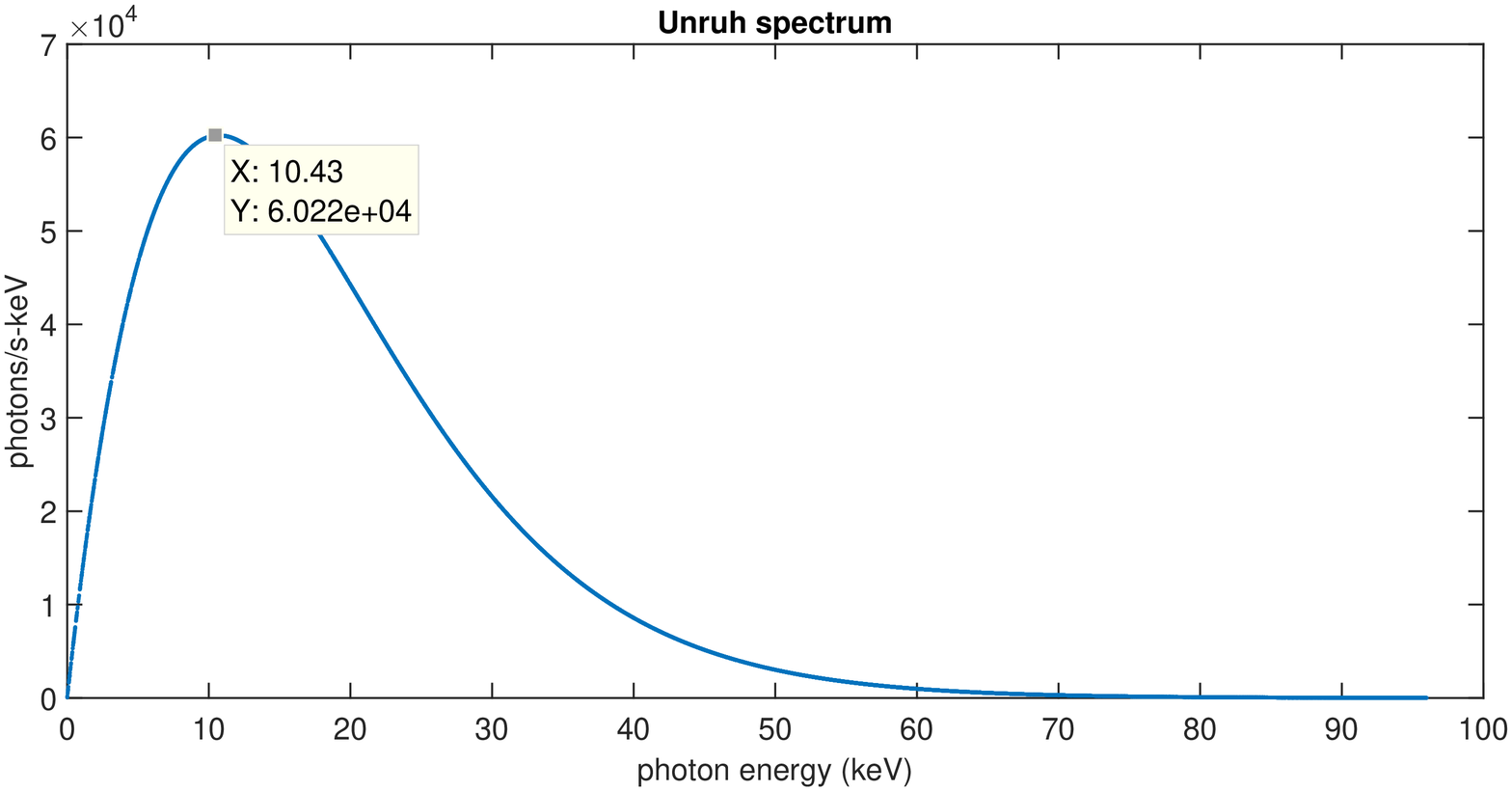}
\caption{Spectrum of scattered Unruh photons 
emitted from the LCLS, as seen in the lab frame.
Some of the virtual Unruh photons present in the electron rest frame  
have scattered from the beam electrons and become real.}  
\end{figure}

\begin{figure}[H]
\centering
\includegraphics[width=130mm,height=80mm]{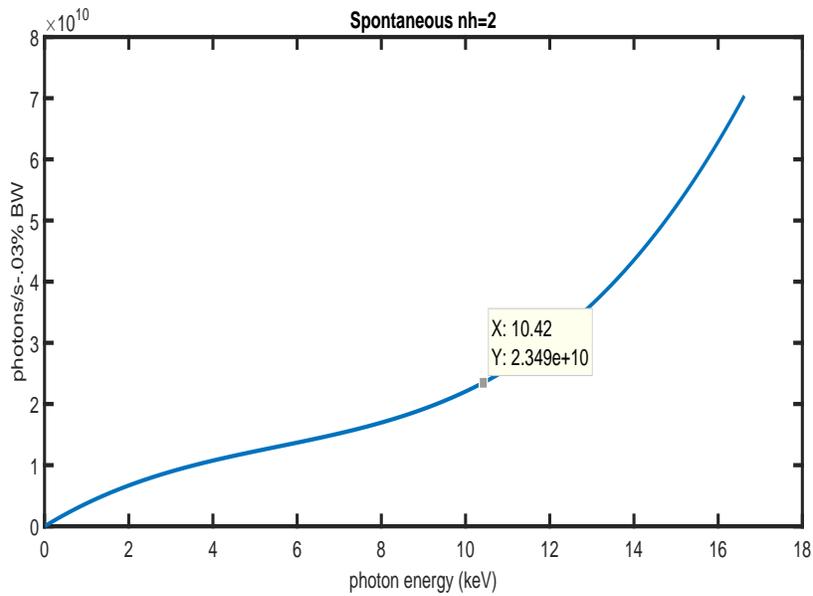}
\caption{The spontaneous radiation spectrum at the LCLS, in the region 
of the peak of the Unruh spectrum. The contribution of the 2nd harmonic
is shown.}
\end{figure}

\begin{figure}[H]
\centering
\includegraphics[width=130mm,height=80mm]{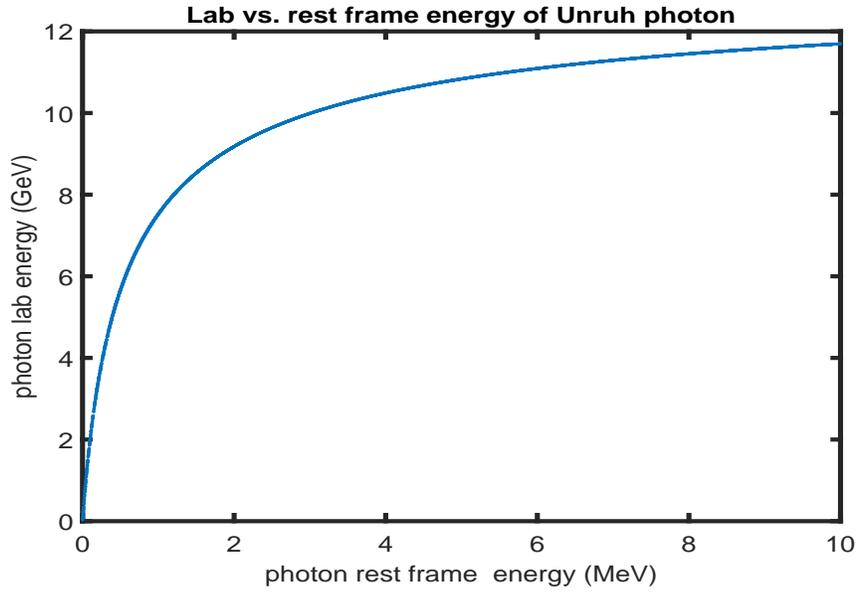}
\caption{Extreme acceleration: the Unruh photon energy in GeV in the lab frame 
 vs its energy in the rest frame, in MeV; for backward incidence
in the rest frame, integrated over scattering angles. Note
that the lab energy never exceeds the energy of the electrons.}  
\end{figure} 

\begin{figure}[H]
\centering
\includegraphics[width=130mm,height=80mm]{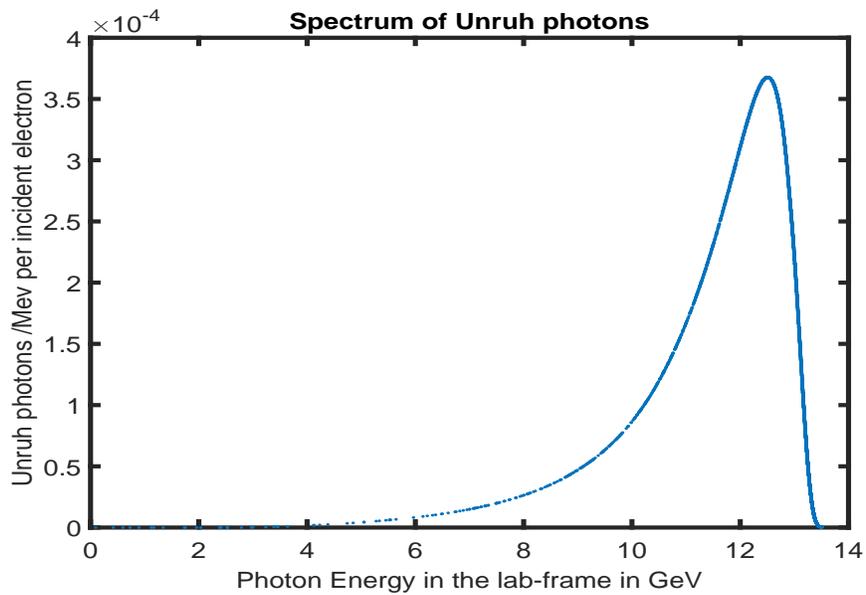}
\caption{Extreme acceleration: spectrum of the Unruh photons per electron per MeV vs their 
energy in GeV in the lab frame.}
\end{figure}

\end{document}